# STAR FORMATION IN THE ERA OF THE THREE GREAT OBSERVATORIES:

Scott J. Wolk[1], Norbert Schulz[2], John Stauffer[3], Nancy Evans[1], Leisa Townsley[4], Tom Megeath[1], Dave Huenemoerder[2], Claus Leitherer[5], Ray Jayawardana[6]

[1] Harvard-Smithsonian Center for Astrophysics, 60 Garden St., Cambridge MA, 02138 USA
[2] Massachusetts Institute of Technology, 70 Vassar St., Cambridge, MA 02139, USA
[3] Spitzer Science Center, California Institute of Technology, 1200 East California Boulevard, Pasadena, CA 91125 USA
[4] Department of Astronomy and Astrophysics, 525 Davey Laboratory, Pennsylvania State University, University Park, PA 16802 USA
[5] Space Telescope Science Institute, 3700 San Martin Drive, Baltimore, MD 21218 USA
[6] Department of Astronomy & Astrophysics, University of Toronto, Toronto, ON M5S 3H8, CANADA

## ABSTRACT

This paper summarizes contributions and suggestions as presented at the *Chandra* Workshop *Star Formation in the Era of Three Great Observatories* conducted in July 2005. One of the declared goals of the workshop was to raise recognition within the star formation research community about the sensible future utilization of the space observatories *Spitzer*, *Hubble*, and *Chandra* in their remaining years of operation to tackle imminent questions of our understanding of stellar formation and the early evolution of stars. A white paper was generated to support the continuous and simultaneous usage of observatory time for star formation research. The contents of this paper have been presented and discussed at several other meetings during the course of 2005 and January 2006.

Subject headings: General --- meeting summary, Stars --- formation, ISM --- general

## 1. INTRODUCTION

As part of *Chandra's* ongoing series, a workshop was held on 13-15 July 2005, entitled: *Star Formation in the Era of Three Great Observatories* (***http://cxc.harvard.edu/stars05***). One goal of the workshop, which was co—sponsored by the *Spitzer* Science Center, was to develop a "white paper" to serve as a roadmap for the study of star formation from space. We sought to review topics in star-formation which are inherently multiwavelength, and define both the current state of our knowledge and the points of current controversy where new observations are most needed. Because of this dual role, we include here some material which was not discussed at the meeting itself, but came out of discussions at other meetings where this white paper was presented. These meeting include the "X-ray Universe" held in El Escorial Spain 25-30 September 2005, "Protostars and Planets V", held in Waikaloa, Hawaii 23-28 October 2005 and "Six years of Chandra" held in Cambridge Massachusetts 2-4 November 2005. Slides from the individual talks given during the July 2005 workshop are available at http://cxc.harvard.edu/stars05/agenda/program.html and are referenced in ADS astronomy abstract service.

At the beginning of the workshop, Lee Hartmann identified three periods of star formation which are specifically poorly understood. These included the formation of protostellar cores, disk accretion and planet formation. The science organizing committee (SOC) of the meeting also identified key areas which highlight the complementary aspects of these observatories, which can be summarized as stellar populations, the formation and evolution of disk systems, and rotation and dynamos. From the standpoint of the star formation process, Thierry Montmerle pointed out at the end of the workshop, that the wavelength bands in which the three space observatories operate have each a distinct significance, *Spitzer* in the infra-red band for dust, *Hubble* in the optical band for gas, and *Chandra* for the X-ray band for magnetic activity, with the understanding that there are topical overlaps in all the bands. In this respect, we focused on these subtopics for which *Spitzer, Hubble*, and *Chandra* have the most to contribute during this unique period of operation. We also considered observations from other facilities including radio and other ground based facilities at all available wavelength windows. Emphasis was also placed on theoretical work to support these observations.

## 2. CAPABILITIES OF THE GREAT OBSERVATORIES

The Great Observatories see structure (HST), thermal states and embedded stars (*Spitzer*) and hot gas and high energy sources (*Chandra*) within the ISM as well. Three speakers outlined the capabilities of these instruments in the context of star formation studies. *Hubble* traces the evolution of the ionized gas. In its first 15 years HST has changed how we view almost all aspects of the latter phase of star formation. Figure 1 illustrates some of the key capabilities of the Great Observatories. The following sections list and comment on presentations that focus on the capability of each observatory separately as well as in combination.

**2.1 Individual Accomplishments**
*Spitzer* traces dust. In a summary presentation, L. Allen asserted that *Spitzer* has the unique capability to study dusty environments including the evolution of dust. Recent work from Allen et al. (2004), Megeath et al. (2004) and Muzerolle et al. (2004) have demonstrated the ability of IRAC and IRS to discern stars with disks and stars with infalling envelopes (protostars) from both each other and diskless stars (out to some limit) with ease. Tracing the evolution of the ionized gas with *Hubble* was summarized by D. Padgett. Several highlights were identified, which include the structure and population of massive star formation regions including the evaporation of molecular clouds by O stars, the discovery of "Proplyds" and outflows and accretion. The latter also emphasizes the imaging of Herbig-Haro objects. Magnetic activity is a trademark of *Chandra* studies. Highlights of the first six years of *Chandra* were presented by S. Wolk. *Chandra* has demonstrated that X-rays dominate cosmic rays as a source of ionization near PMS stars. The detection of complex flares in pre-main sequence stars and possibly protostars was discussed as well as the detection of X-ray fluorescence by protostellar disks during giant flares.

**2.2 Demonstrations of Combined Observations.**
There were several presentations which demonstrated the power of combining data from the Great Observatories. The most striking work was presented by B. Brandl who showed vast regions of 30 Doradus and other super clusters which appeared void and empty to *Spitzer* are seen by *Chandra* as filled with plasma with temperatures between 5 and 20 MK. The cavities are off center relative to the current epoch of star formation implying that X-ray emission traces the *previous* generation of O-stars, at least in 30 Doradus. B. Whitmore presented results for the Antennae galaxies showing similar interplay of plasma and warm dust on a galactic scale. J. Forbrich presented results of radio sources in the Coronet and found IRS 5 was the most variable source in both X-ray and cm wavelengths, but lacking simultaneous data, it was difficult to understand the physical connection.

A full session was spent on star formation throughout the constellation of Orion. Figure 2 shows the ONC as view from Chandra, HST and in the infrared. E. Feigelson outlined results from 13 papers in a special ApJ supplement devoted to the Chandra Orion Ultradeep Project (COUP). Several key results came from this study, notably that observation that Solar-type stars exhibit their highest levels of magnetic activity during their PMS phases, X-ray ionization will dominate cosmic ray ionization of molecular cloud cores if a stellar cluster is present, X-rays can efficiently irradiate proto-planetary disks, X-rays dominate disk ionization and may alter disk structure, dynamics & chemistry, if MHD turbulence is induced, planet formation processes may be substantially affected. The X-ray data also support models of particle irradiation of meteoritic solids. T. Megeath then presented very early *Spitzer* results from the same region. He showed that while HST was able to image the outer regions (900 AU) of some disks, spectral energy distribution measurements made with IRAC could detect structures as close in as 10 AU. He also showed that IRAC was sensitive to COUP X-ray sources with disks. In fact, when sources were limited to those with detections in all four IRAC bands, the results were dominated by X-ray sources with disks.

## 3. OPEN QUESTIONS
Here we summarize the open issues raised during the meeting as well as in discussions at special splinter sessions at the end of the workshop. Some items also came up in discussions at succeeding meetings. The session about populations was led by F. Walter, about disks by N. Calvet, and rotation by S. Strom and S. Wolff. Hot star science was discussed in the plenary session, especially in the talks given by B. Brandl and B. Whitmore. Barbosa & Figer (2004) recently compiled a list of about 120 key questions on the topic drawn from 12 contributors. Neither this work, nor the broader topic of hot star evolution, was revisited in detail.

**3.1 Populations**
All three telescopes have made significant contributions to our understanding of populations in young clusters. *HST* excels at binary studies and objects at the bottom of the IMF. *Spitzer* is crucial to detecting very young and low mass objects, as well as detecting the cooler disks. *Chandra* works orthogonally to *Spitzer*, identifying objects as young independent of the disk characteristics. But *Chandra's* X-ray sensitivity can be a source of confusion in the definition of a young stellar object. Since X-rays persist at some level throughout a low mass star's lifetime zero-age main sequence stars (and older stars) can be detected in deep X-ray observations. Five key questions have emerged from the discussion.

*What is the population structure of a star forming association?* The scope of this question ranges from the definition of an initial mass function of stellar clusters, its dependence on cluster location, environment, initial cloud properties as well as star formation history. Within a given box on the sky we can find cluster members spanning 10 Myr in age. To construct the IMF for a cluster we need to determine the luminosity. To do this, we have to correction account for completeness, distance, absorption, and disks. This requires a multi-wavelength approach (X-ray, optical, near-IR and mid-IR) to identify complete populations from embedded objects to naked stars.

*What are the main drivers in the evolution of protostellar disks?* The identification of the general sequence of events by which a system evolves from hosting a full optically thick disk to a presumably naked star without disk signatures is still rather incomplete. The standard framework (Shu, Adams & Lizano 1987) still has general acceptance, but it is not clear if a star has to pass through all phases, or which range of time scales are involved. One related question involves the issue of disk efficiency, specifically, if disk accretion could be efficient enough not to leave any material behind.

*How do high-mass star formation and the large amount of plasma generated affect the ISM?* As shown by B. Brandel the plasma is often found adjacent to warm dust. In the case of the LMC, the plasma seems to be a remnant of a previous generation of star formation and is not directly interacting with the active star forming regions of 30 Doradus and R 136. Within our own galaxy, star formation and hot plasma has been observed coincident in RCW 38 and the Rosette.

*How does the X-ray emitting process differ in PMS stars as compared to MS stars?* This question has now gained specific recognition in recent X-ray studies of nearby star forming regions, at the forefront are those of the ONC. Though some major characteristics and differences have been identified, most issues are unresolved. Issues range from rotational activity, star-disk interactions, and high coronal temperatures to the ultimate processes that lead to various levels of magnetic activity. Support has to come specifically from theory about the internal structure of PMS stars and possible early versions of dynamo activity.

*Under which conditions do brown dwarfs form?* Simulations indicate that in multiple formation scenarios should occur. In some cases, such as ejection, these mechanisms truncate growth prematurely. To date, there is not a lot of evidence for a strong mass-velocity relation which would be suggested if all brown dwarfs formed via ejection.

**3.2 Questions about Disks**
*Hubble* has made significant contributions here with the direct detection of disks, e.g. GG Tau, TW Hya and proplyds. *Hubble* validated the flared disk model (HH30) and demonstrated that the angular momentum of stars in a binary system are not necessarily co-aligned (HK Tau/c). *Spitzer* is extremely sensitive to dust emission of a few 100K and is identifying many new targets for HST disk imaging. The detection of X-ray fluorescence of disks by *Chandra* opens the possibility of reverberation mapping of disks on scales < 1 AU by future X-ray spectroscopy missions.

*What is the relation between ionization and accretion in protostellar disks?* The highly ionizing environment in protostellar systems impacts the disk in many ways. Not only may it provide the main means of angular momentum transport and thus fuel the motor for matter to be accreted, these processes may even engage in complex feedback mechanisms between ionization and accretion. Issues also include effects of jet and wind production as well as the possible destruction of the disk.

*Can X-ray ionization maintain disk turbulence?* This question is somewhat related to the one above and is key for the understanding of time scales of post-protostellar accretion, specifically why and when accretion ceases. Maintenance of various levels of disk turbulence may contribute to answering how and when to onset of planet formation occurs.

*What generates the inner disk clearing in transitional disks?* This question addresses the physical situation at the onset of planet formation. The issues here span from the possibility of repeated episodes of inner disk clearings to the processes that eventually leads to dust settling and the growth of planetesimals. Finally the role of rotation needs to be investigated.

### 3.3 Questions about Rotation

Rotation periods and *v*sin*i* are generally studied from the ground. But rotational modulation has now been directly observed in the X-rays (E. Flaccomio). The physics manifest themselves in multiple ways. One of the strongest of these is the X-ray flux-rotation correlation which was very strong in the older clusters observed with ROSAT. These relations are not seen in the younger clusters favored for observation by *Chandra*. If disk-coupling is critical to the bimodal rotation distribution seen in some cluster, *Spitzer* should be able to measure the disk lifetime statistics and learn if they are consistent with the field rotation distribution. Early results from a study by L. Rebull finds that excesses do not necessarily imply longer periods, but longer period stars are more likely than shorter period rotators to have mid-IR excess.

*How can we understand the rotation-activity correlation, or the lack of, in PMS stars?* Recent analyses in the ONC (Feigelson et al. 2002, Flaccomio et al. 2003), of radiative activity correlated with rotational periods showed an uncharacteristic lack thereof for stars of ages ~1 Myr. This possibly implies that coronal activity and dynamos differ fundamentally with respect to MS stars. Specifically, a possible decrease in X-ray activity at high rotation rates may suggest some sort of super-saturation.

*Are there various PMS populations that distinguish themselves with respect to X-ray activity?* Recent studies of X-ray activity in Orion suggested the possibility of a very slowly rotating population at age ~1 Myr, which is relatively X-ray inactive.

*Is there a change in the rotational velocity distribution in PMS stars above or below some stellar mass?* The correlation of rotation and mass in late type PMS stars is poorly understood esspecially for the lower mass range.

*How does the rotational velocity distribution evolve from the PMS to the MS phase?* Rotational velocity distributions of low mass post-PMS stars in the Pleiades and $\alpha$ Persei, for example, need to be compared with the ones observed during the PMS and MS phase.

### 3.4 Questions about Diffuse X-ray Emission and Dust Structure

*Chandra* has been a powerful tool in understanding the lifecycle of matter. It is the only instrument which can unambiguously identify the diffuse plasma present in massive star forming regions. In this sense, we are getting our first look at the driving forces in the cycle of stellar evolution. It is most likely from the *Spitzer* and other mid-IR data we will be able to understand the dominant energetics of these systems, whether the plasma triggers star formation or the dust contains the plasma bubbles. Figure 3 shows one of many examples of close interaction between plasma seen in X-ray emission and dust seen at mid-infrared wavelengths, similar examples have been seen in with optical forbidden line emission.

*Where do the SNe that produce the X-ray superbubbles come from?*
We should be able to identify the previous generation of star formation which included the formation the progenitors to these supernovae.

*Why are many massive clusters sitting at the edges of dense clouds?* Whether this is a selection effect of some sort or not, the birth of massive clusters as well as their subsequent interaction with its remnant natal cloud as complex and far from understood.

*Why do most massive star-forming GMCs look so similar?*

*What is the sub-structure and content of extragalactic IR super star clusters?*

### 4.  Key Observations and Resources

In order to answer the open questions outlined above, we collected a list of current and proposed observations. It is the synthesis of various individual observations that allow us to address more fundamental questions of physical processes involved. This list should serve as both information and inspiration to the community about ongoing and planned research in observational star formation research. We also tried to assess the potential of future resources to continue this quest.

**4.1 The Large Magellanic Cloud as a Laboratory**
The group was reminded by You-Hua Chu that the LMC and SMC make good laboratories for many experiments since extinction is low and the issues regarding the relative luminosity of objects are much diluted. Since many objects of interest appear nearly face-on, structures relating to triggered star formation can be directly identified. Theoretically, individual high mass stars can be resolved, however this is complicated by the fact that most high mass stars reside in young clusters. Finally, there is a publicly available survey of the LMC which has been performed by *Spitzer*. A matching survey by *Chandra* is worth consideration.

**4.2 Theory**
To be clear, the conference focused on recent observational results and future goals. New theoretical work was not the primary concern except to the extent that it could direct future observations. As pointed out in the introductory talk (L. Hartman), while most of us are working within the outline presented by Shu, Adams & Lizano (1987), the physics of the individual phases is not well understood (with the exception of the formation of the flattened disk structure). One suggestion was to focus of the first step by improving our understanding of the formation and evolution of the magnetic field and the temperature and pressure balance of the ISM. This is a dividing topic in the community - ISM turbulence versus the standard magnetic paradigm of star formation, has top be addressed. It is not only "improving" but also resolving an apparent divide in the community, where both theory and observations are needed to reconcile the views. While likely both views have their place in our understanding of the initial onset of star formation, research has to focus on where that place is and how a concerted view improves our interpretations. This is clearly a field where observations and theory directly interface in terms of the evaluation of time scales, magnetic fields, ionization fractions, shocks etc. The second suggestion also focused on the earliest phases, to develop numerical simulation and visualization codes of sophistication sufficient to span the ranges of temperatures, densities and chemical conditions that occur during the epoch when protostars emerge from molecular cloud complexes.

**4.3 Current Space Observatories**
Our focus is on what observations need to be made now, while *Spitzer*, *Hubble* and *Chandra* still operate simultaneously operating and can act one the results of each other. *Spitzer* has the lead role in identifying embedded young stellar objects and objects surround by cool dust. IRAC and MIPS photometry is needed to classify low-mass YSOs. (Some examples were shown by S. Carey, J. Muzzarole, R. Gutermuth and L. Cambresy). *Spitzer* needs to provide a catalog of embedded protostellar objects just emerging from the envelope infall phase to fully revealed star/disk systems for follow up by HST and other telescopes. *Chandra* has observed a remarkably large fraction of clusters within 1 kpc, >25% of the clusters and >60% of the stars in the Porras et al. (2003) sample. The median depth of these observations ($L_X$~28.3) is sufficient to detect all stars above $1M_\odot$ and 90% of all stars above 0.3 $M_\odot$ (Feigelson et al. 2005). This is a good start to complimenting the *Spitzer* cores to disks (C2D) program. The Formation and Evolution of Planetary Systems (FEPS) program focuses on older, more isolated stars and is not well surveyed by the current generation of X-ray telescopes. The goal should be spatially complete X-ray maps to a constant luminosity limit for all the C2D and FEPS fields. An X-ray counterpart to the GLIMPSE survey also had support. There is a dearth of massive revealed clusters between 0.5 (ONC) and 1.5 kpc (M16, M17, M20, Tr 14 Tr16, RCW 38 and the Rosette are all between 1.7 and 3 kpc). Typical 100ks *Chandra* observations reach log $L_x$~30 for clusters at 2 kpc which is enough to detect about half of the solar mass stars. Deeper *Chandra* observations are needed to identify cluster populations. Currently approved future X-ray observatories will not resolve distant, young galactic cluster stars. At 2 kpc and an $A_V$ ~ 4, a 600 ks ACIS-I exposure is needed to achieve a 2-8 keV log $L_x$ ~ 28.8 for a cluster of about 1Myr, thereby detecting half of all stellar cluster members (this was specifically suggested by M. Gagné). There is additional interest in the role of instabilities and/or turbulence in stabilizing planetary orbits. This could be pursued through theory grants available through the *Chandra* announcement of opportunity. *HST* is still the ultimate instrument for high resolution spatial imaging. *HST* is needed to carry out pathfinder imaging and spectroscopic observations of proto-stellar envelope morphology and kinematics. It is also the only instrument capable of direct imaging of most star disk systems which are indicated by the *Spitzer* data. While adaptive optics can achieve similar results from the ground this work is of primary use in the near and mid IR and still requires bright guide stars. Thus, a very limited amount of the sky is available via AO.

**4.4 Future Space Observatories**
The time and motion domains are still poorly explored. Space/time (astrometry, 3D), and Doppler surveys are needed to understand the long term evolution of clusters. Spatial and spectral capabilities in the mid-IR lag behind the optical and near-IR. The far-IR and sub-mm regimes are still barely explored. These areas are critical to understanding the earliest phases of star formation and the evolution of thick disks. Below we list, in approximate order of availability, important experiments for upcoming space observatories. The meeting participants avoided a strict prioritization of the various missions; instead concerns about each mission were discussed in addition to the science goals.

**HST Servicing** – The continuing viability of HST is critical to studies of star formation, especially in low mass clusters and more distant clusters where guide stars for AO are not guaranteed. The science lost if HST is scuttled would be difficult to perform using other observatories and instruments. Although we do not see a clear alternative to the HST servicing mission being offered by NASA, it is also realized that slipping or even cancellation of other programs is as crucial. The fact that there is not much choice is a painful reminder of the reality that available options are limited. In that respect though there was no sensible assessment of the priority of HST servicing by the group with respect to the other observatories, such a mission is clearly supported from the science perspective.

**Ultraviolet Spectroscopy** – With the loss of HST/STIS and constraints on FUSE there is capability to perform high resolution spectroscopy in the range between 1100Å-3000Å is somewhat limited. This is the key wavelength range for the study of accretion in pre-main sequence stars and to study the ISM near hot massive stars often found at the centers of young clusters. The Cosmic Origins Spectrograph (COS) scheduled for the HST servicing mission is intended to fill this void. However COS is not designed as a replacement for STIS, it was built for spectroscopy of faint sources in uncrowded fields. We are encouraged that NASA is studying the possibility of repairing STIS if a servicing mission occurs. Still, a concern within the community is both current lack of a UV spectrograph and the lack of this capability in the foreseen future.

**SOFIA** – There is every expectation that SOFIA can have the type of fundamental scientific impact that is usually associated with Great Observatories well into the post-Spitzer, post-Herschel era, provided proper investments are made in future state-of-the-art instruments (Werner et al. 2004). SOFIA provides the only planned far-IR imaging and spectroscopic capabilities after Spitzer, with three times better spatial resolution. SOFIA will continue sub-millimeter spectroscopic capability after the end of Herschel's 3-year mission. The mission is zeroed in the FY 2007 budget due to a combination of project delays and NASA budget difficulties. If that decision is not reversed there would be two major holes in the available wavelength coverage. A key science goal of the SOFIA mission includes the detection and measurement of tracers of shocked and radiatively heated gas produced in infalling envelopes, at envelope/accretion disk interfaces and by accretion-disk-driven winds and jets.

**Herschel (~2008)** – One of the primary science goals of the far-IR imaging and spectroscopic observatory will be to survey cold cores within 500 pc of the Sun. *Herschel* will be able to measure the SEDS of massive cold cores. These data can be correlated with X-ray emission from the interstellar medium to understand ISM absorption. Herschel will observe the cold phase ISM. The interaction of this with the warm dust ISM seen by *Spitzer* is unknown. *Herschel* will be uniquely suited to look for the cool disks around substellar objects.

**WISE (~2009)** – The Wide-field Infrared Survey Explorer is not primarily designed as a star formation instrument. It should be more sensitive than SOFIA and will cover wavelengths which Herschel will not cover. The ~5" resolution will limit it somewhat, but this will be fabulous project for identifying every site of star formation in the nearest 1 kpc. Some follow up will be possible with SOFIA and Herschel, but JWST will be needed to follow up the observations with higher angular resolution and equivalent or better sensitivity at 3-20 μm.

**GAIA (~2012)** – There was some concern that the data from this mission might be made public very slowly. This concern comes from the lack of NASA involvement and the lack of a limited proprietary period. These factors couple with the experience of HIPPARCHOS and ROSAT in which the datasets took nearly a decade to become public. In addition, there are ground based programs such as PANSTARRS and LSST which should resolve distances to many clusters within 200-300 pc using trigonometric parallax to a few, carefully studied stars in each cluster. There will still be important questions for GAIA, most notably its primary mission – to determine the distance to many clusters in the 300-500 pc range, especially the various parts of the Orion Complex. GAIA will also obtain optical photometry to well below the brown dwarf limit for all clusters within 1 kpc although identification of cluster members at faint magnitudes will still be a challenge due to background confusion. To help identify cluster members, GAIA will measure the cadence and magnitude of the optical variability of PMS stars as well as measuring the change as a function of time. This requires extensive monitoring of a range of clusters.

**JWST (~2013)** – JWST will be a near and mid infrared instrument of unsurpassed resolution and sensitivity. Given high resolution spectroscopy, it could quantify envelope infall rates for a large sample of forming stars spanning a wide range in mass but *spectroscopic capability beyond R~3000 is not planned for baseline mission.* It is not clear if this decision is final, however given the recent discovery of the "undercosting" of the mission, it is doubtful that additional scientific capability can be added. JWST *will* be able to quantify the shape of the stellar initial mass function (IMF) to masses as small as 10 Jupiter masses in nearby star-forming regions. JWST will be so sensitive it will be able to quantify the IMF down to the hydrogen burning limit in local group galaxies in regions spanning a range of metal abundances.

**Constellation-X (~2017)** – There was a lot of discussion at the meeting about the resolution of the gratings proposed for Con-X. The view of the community was that X-ray studies in star formation need at least moderate resolution. The current goal of the mission (resolution of 3000) is consistent with that requirement. The requirements for the calorimeter are 4eV at 6keV (R=1500) and 2eV at 1keV (R=500). The grating requirements are not as well specified but must meet the observatory requirement of R>300. Design studies with the gratings are ongoing and detailed on the Con-X website. Studies of technologies such as off-plane reflection gratings are under way and the community welcomes such a development. Assuming resolution of about 1500 at 1 keV and a microcalorimeter, primary goals include: **1)** Survey nearby TTS for accretion and infall signatures in their forbidden lines. **2)** Survey nearby TTS using reverberation mapping to map the surface structure of their disks. **3)** Survey all stars within 10 pc for cometary emission as evidenced by charge-exchange.

**Other X-ray Missions** – A major concern about Constellation-X is that the spatial resolution is significantly poorer than that of Chandra and that the time frame was quite prolonged. One idea put forward was a lower cost multilayer telescope. Such a telescope may be able to exceed 0.1" resolution at energies well below 1 keV. Such a telescope could do very interesting work on multiple systems in Taurus, the Hyades and the TW Hya association. Further, there is the possibility of observing charge exchange from comets orbiting stars within 10 pc of the Sun. However, since such optics would be confined to the soft X-ray band it would not be able to continue the work of Chandra in regions of massive and embedded star formation. This would require active grazing incidence optics such as that discussed by the Generation-X vision mission.

**ALMA** – While the meeting focused on space based observatories there was discussion of both sub-millimeter and millimeter observing. The SMA is a forerunner of ALMA, Q. Zhang presented early results from the SMA which has already demonstrated its ability to detect and study cold dust and gas at arcsecond resolution which translates to 100 AU physical scales. Further, the lines it accesses allow us to probe different physical regimes of disks, and jets. ALMA will lower the physical scales by a factor of 5-10 along with an increase in sensitivity. This will allow ALMA to probing the depletion zones of starless cores, measure the proper motion of jets at high time cadence and perhaps even directly image of disk gaps.

**Funding Opportunities –** The NASA Great Observatories have a reasonable mechanism for multi-wavelength programs. However, most ground based observatories generally require a wavelength focused approach and are not designed to support synthesis of observations. This emphasizes the importance of the NASA/ADP and possible "(N)VO"-like opportunities. There general consensus was that NOAO and the NASA great observatories do a good job of enabling multiwavelength observations. Extensive discussion on optimizing utilization of those resources followed but did not lead to specific recommendations. There was support for "collaborations of scale" sufficient to develop the numerical codes and visualization tools necessary to model the physical and chemical evolution of molecular clouds. NSF is the most natural avenue for this.

## 5. Summary

We were very gratified by the focus and high level of discussion among the ~120 participants. The workshop was very short, only three days, and concentrated on recent results, especially from *Chandra* and *Spitzer*. Only one morning was devoted to problems in the field and little time was left to explore solutions. The goal of generating a white paper with detailed concepts is, of course, quite ambitious and we are aware that much is still left to formulate. In this respect the paper generated from such a workshop can only be the beginning of the necessary discussion about the future of star formation research from space.

Acknowledgements: We would like to thank the CXC director Harvey Tananbaum, for agreeing to make this workshop a reality and for his active participation in the meeting discussion. We would also like to thank George Helou, executive director of IPAC, and B. Thomas Soifer, Spitzer Science Center director, for their support of this meeting. Finally we thank the attendees, without whom, none of this discussion would have ever occurred.

**References:**

Allen, L. E., et al., 2004, ApJS, 154, 363
Barbosa, C., Figer, D., 2004, astro, arXiv:astro-ph/0408491
Brandl, B.R., et al., 2005, IAUS, 227, 311
Feigelson, E. D., et al., 2005, ApJS, 160, 379
Feigelson, E. D., Broos, P., Gaffney, J. A., Garmire, G., Hillenbrand, L. A., Pravdo, S.H., Townsley, L., & Tsuboi, Y. 2002, ApJ, 574, 258
Flaccomio, E., Damiani, F., Micela, G., Sciortino, S., Harnden, F. R., Murray, S. S., & Wolk, S.J. 2003, ApJ, 582, 398
Megeath, S.T., et al., 2004, ApJS, 154, 367
Muzerolle, J., et al., 2004, ApJS, 154, 379
Porras, A., Christopher, M., Allen, L., Di Francesco, J., Megeath, S.T., Myers, P.C., 2003, AJ, 126, 1916
Shu, F.H., Adams, F.C., Lizano, S., 1987, ARA&A, 25, 23
Werner, M. (chair) 2004, Independent Science Operations Review (ISOR) report on SOFIA

# Figures:

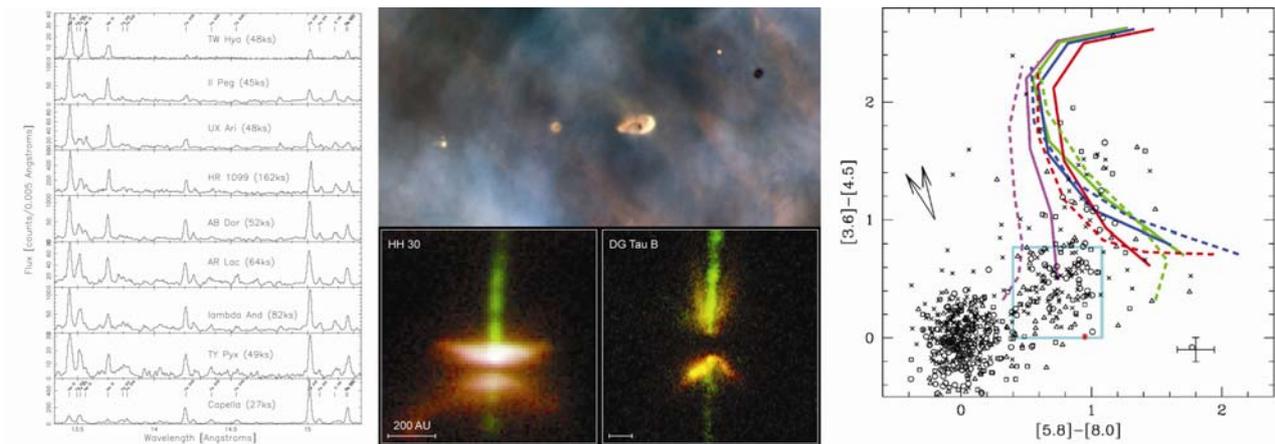

**Figure 1:** Key Capabilities: The current generation of X-ray spacecraft features spectral resolution of R~1000. This allows sensitive diagnostics of the conditions under which X-rays are generated. All 3 of the great observatories have good spatial resolution allowing them to explore the morphology of individual objects. *Hubble*, with the best resolution, has proven especially successful in disk studies. The *Spitzer* IRAC bands make it very sensitive to the identification of disk systems.

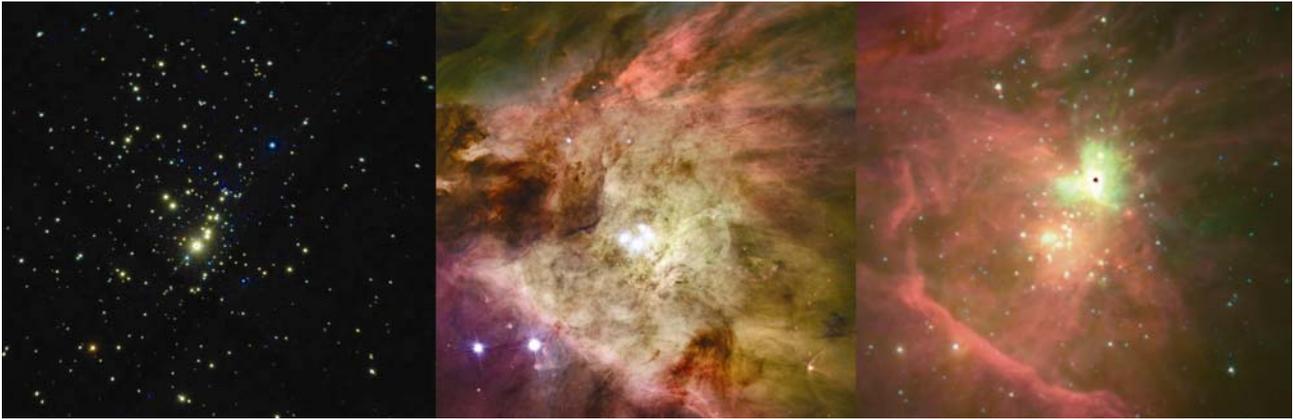

**Figure 2:** Orion as an example: The images above show X-ray (*Chandra*), optical (*HST*) and mid-IR(*Spitzer*) views of the central 7' of the ONC. The near-IR and X-ray images trace the stellar population equally well. The IR-luminosity traces the bolometric luminosity; the X-ray luminosity traces the magnetic fields or winds. This optical/HST image is dominated by the gas. The Spitzer image shows strong diffuse PAH emission the Kleinman-Low Nebula is overexposed.

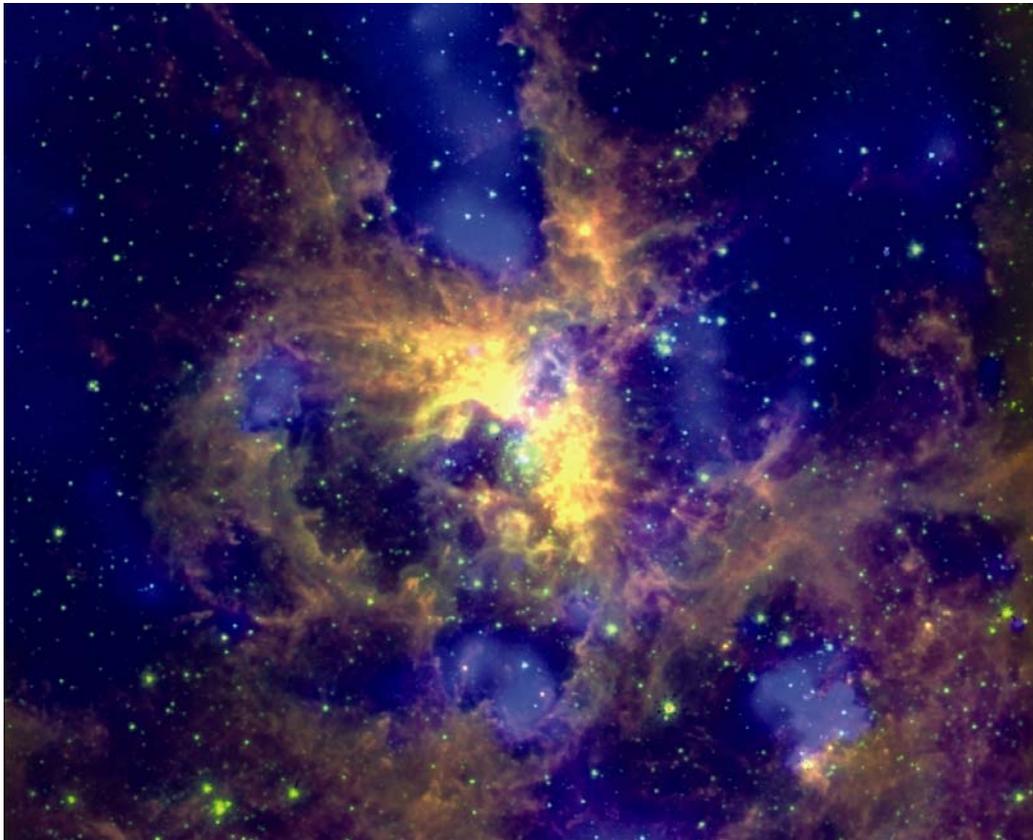

**Figure 3:** The central 200pc of 30 Doradus as imaged by *Chandra* ACIS (blue: 500-700eV) and *Spitzer* IRAC (green: 3.2-4.0 μm, red: 6.5-9.4 μm) (Brandl et al. 2005).